\documentclass{easychair}

\usepackage{amssymb}
\usepackage{graphicx}
\usepackage{wrapfig}
\usepackage{subfloat}
\usepackage{pifont}
\usepackage{float}
\usepackage{url}
\usepackage{tabularx}
\usepackage{subfig}

\newcolumntype{C}[1]{>{\centering\arraybackslash}p{#1}} 
\newcolumntype{R}[1]{>{\raggedleft\arraybackslash}p{#1}} 

\newcommand{\saucy}{{\bf saucy}}
\newcommand{\nauty}{{\bf nauty}}
\newcommand{\bliss}{{\bf bliss}}
\newcommand{\traces}{{\bf traces}}
\newcommand{\nishe}{{\bf nishe}}

\newtheorem{mydef}{Definition}

\begin{document}

\title{Graph Symmetry Detection and Canonical Labeling: Differences and Synergies}

\titlerunning{Symmetry Detection and Canonical Labeling: Differences and Synergies}

%
%
\author{Hadi Katebi \and Karem A. Sakallah \and Igor L. Markov}
%

\institute{EECS Department, University of Michigan\\
\{hadik,karem,imarkov\}@umich.edu}

\authorrunning{Katebi, Sakallah and Markov}
\clearpage 
%
%

\maketitle

\begin{abstract}
Symmetries of combinatorial objects are known to complicate search algorithms, but such obstacles can often be removed by detecting symmetries early and discarding symmetric subproblems. Canonical labeling of combinatorial objects facilitates easy equivalence checking through quick matching. All existing canonical labeling software also finds symmetries, but the fastest symmetry-finding software does not perform canonical labeling. In this work, we contrast the two problems and dissect typical algorithms to identify their similarities and differences. We then develop a novel approach to canonical labeling where symmetries are found first and then used to speed up the canonical labeling algorithms. Empirical results show that this approach outperforms state-of-the-art canonical labelers.
\end{abstract}

\section{Introduction}
\label{sec:intro}

Combinatorial objects and structures in many applications are often represented by graphs. Examples of such structures appear in domains like computer networks \cite{Lancichinetti-2009,Cheswick00,Govindan00}, electronic circuits \cite{Velev-DAC02,ISPD2005}, and even mathematical chemistry \cite{bonchev1991chemical,Balaban-2005}. The challenges that arise in those areas are usually addressed by graph-theoretical problems. Two of those problems, with major industrial significance, are graph \emph{symmetry (automorphism) detection} and \emph{canonical labeling}. A symmetry is a permutation of the graph's vertices that preserves the graph's edge relation. A canonical labeling is a labeling of the graph's vertices that uniquely captures the structure of the graph, and serves as a signature that is invariant under all labelings. In a common application, the symmetries of molecules can be used to predict and explain chemical properties \cite{hargittai2009symmetry}, and the canonical labeling of chemical compounds can help build a database of chemicals \cite{RandicBW81}. 


Graph canonicalization and automorphism detection share similarities in both theory and implementation. These two problems are poly-time equivalent; however, since they are not believed solvable in worst-case polynomial time (unless P=NP), this says little about practical algorithms for these problems. Given that these problems are also not believed to be NP-complete (again, assuming P$\neq$NP), there is no {\em a priori} need to use the full computational machinery developed within high-performance algorithms for Boolean satisfiability.

Although symmetry detection, canonical labeling and their poly-time equivalent problems, such as \emph{graph isomorphism} (two graphs are \emph{isomorphic} if one can be mapped to the other by a bijection on vertices), are not generally known to be in P or NP-complete, they can be solved in polynomial time for special cases and with high probability for random graphs. Such special cases include bounded-degree graphs \cite{Luks_1982}. Also, an approach from \cite{Babai-79} solves canonical labeling in linear time on average with the probability of $1-\exp(-cn\; \log\: n/\log\: \log\: n)$ (for $n$ vertices and constant $c>0$). For more information on graph isomorphism and its complexity, see \cite{kobler-93}.

Another remark is that the symmetries of a graph map each labeling to another labeling. Therefore, if all symmetries are known, it may be sufficient to visit only one labeling from each equivalence class. This is why all existing graph canonicalization tools also find automorphisms along the way during the search.

The first practical tool to canonically label graphs was \nauty{} \cite{nautyUG,McKay81}, developed by McKay in 1981. The \nauty{} search tree is based on the branching and backtracking framework, which is optimized by integrating \emph{group-theoretical}\footnote{Group theory is a branch of abstract algebra that studies the algebraic structures known as \emph{groups}. A group comprises a non-empty set of elements with a binary operation that is \emph{associative}, admits an \emph{identity} element, and is \emph{invertible}. For example, the set of integers with addition forms a group.} techniques. The involvement of group theory in the search for symmetries and a canonical labeling is due to the fact that the set of symmetries of a graph forms a group under functional composition (called the graph's \emph{symmetry group}). Other canonical labeling tools, such as \bliss{} \cite{Junttila07,Junttila-2011}, \traces{} \cite{Piperno-2008}, and \nishe{} \cite{Tener08}, follow \nauty{}'s algorithms rather closely, but are designed to address possible shortcomings of \nauty{}'s search tree. In particular, \bliss{} improves the handling of large and sparse graphs, \traces{} incorporates a breadth-first scan of branching choices at each level to identify and prune futile subtrees, and \nishe{} uses the notion of \emph{guide tree} to make better branching choices.
 
Unlike \nauty{}-based canonical labelers, another software package, called \saucy{} \cite{Darga-DAC04,Darga-DAC08,Katebi-2010}, was optimized to only find graph symmetries. The data structures and algorithms in \saucy{} take advantage of both the sparsity of input graphs and the sparsity of their symmetries to attain scalability. The algorithmic advances in all versions of \saucy{} have separated the search for symmetries from the search for a canonical labeling. The experimental results in \cite{Katebi-2011} confirm that \saucy{} is currently the most scalable symmetry-finding tool available. 

In this work, we highlight the algorithmic differences between the search for symmetries and the search for a canonical labeling. We particularly focus on the algorithms implemented in \saucy{} version 3.0 \cite{saucy-3.0} and \bliss{} version 0.72 \cite{bliss-0.72}, but our comparison (with minor changes) can be extended to other \nauty{}-based canonical labeling tools. Dissecting \saucy{} and \bliss{}, we point out an intrinsic limitation of the canonical labeling tree, and illustrate how that limitation is relaxed in the search for symmetries. Furthermore, we propose a novel graph canonicalization approach that uses \saucy{}'s efficiency in finding symmetries as a pre-processing step for \bliss{}'s canonical labeling algorithms. Our experimental results show that the combination of \saucy{} and \bliss{} results in an exponentially faster canonical labeler for large and sparse graphs.

In the remainder, Section \ref{sec:prelim} reviews the necessary background and relevant preliminaries. Section \ref{sec:compare} compares the algorithms implemented in \saucy{} and \bliss{} for computing graph symmetries and a canonical labeling. Section \ref{sec:can_approach} explains our proposed graph canonicalization approach that integrates \saucy{} symmetry detection routines within \bliss{} canonical labeling procedure. Section \ref{sec:case_study} analyzes the run time complexity of \saucy{} and \bliss{} for an example graph. Section \ref{sec:results} experimentally evaluates our canonicalization approach on a benchmark suite consisting of very large sparse graphs. Finally, Section \ref{sec:conclusion} discusses conclusions. 

\section{Preliminaries}
\label{sec:prelim}

We assume that the reader is familiar with basic notions of group theory, including such concepts as subgroups, cosets, group generators, stabilizer subgroup, orbit partition, etc. We review most of these concepts here, but additional details can be found in standard textbooks on abstract algebra, e.g. \cite{Fraleigh00}. In this paper, we focus on automorphisms and a canonical labeling of an $n$-vertex \emph{undirected colored graph} $G$ with set of vertices $V=\{0,1,...,n-1\}$. A \emph{permutation} of $V$ is a bijection from $V$ to $V$. We will use both tabular and cycle notation to express permutations. Permutation $\gamma$, when applied to graph $G$, permutes $G$'s vertices, and produces graph $G^\gamma$.
\begin{mydef}
An automorphism (a symmetry) of graph $G$ is a permutation $\gamma$ of $G$'s vertices that preserves $G$'s edge relation, i.e., $G^\gamma = G$.
\end{mydef}
Every graph has a trivial symmetry, called the \emph{identity} (denoted by $\iota$), that maps each vertex to itself. Two graphs $G_1$ and $G_2$ are \emph{isomorphic} if and only if there exists permutation $\gamma$ such that $G_1^\gamma =G_2$. 
\begin{mydef}
A canonical labeling of graph $G$ is an isomorphism-invariant labeling of $G$'s vertices, i.e., two graphs $G$ and $H$ have the same canonical labeling if and only if they are isomorphic to each other.
\end{mydef}
The set of symmetries of $G$ forms a \emph{group} under functional composition. This group is called the \emph{symmetry group} of $G$, and is denoted by $Aut(G)$. A \emph{generating set} of $Aut(G)$ is a subset of the symmetries of $Aut(G)$ whose combinations under functional composition generate $Aut(G)$. 
Given graph $G$, a symmetry detection tool looks for a set of generators for the symmetry group of $G$, and a canonical labeling tool seeks a canonical representation for $G$. 

A \emph{subgroup} of $Aut(G)$ is a subset of the elements of $Aut(G)$ that forms a group under functional composition. Given $Aut(G)$, the \emph{stabilizer subgroup} of $i\in V$, denoted by $Aut_i(G)$, is a subgroup of $Aut(G)$ that fixes vertex $i$, i.e., $Aut_i(G)=\{\gamma \in Aut(G) | \gamma i = i \}$. In other words, $Aut_i(G)$ contains symmetries of $G$ that map $i$ to $i$. 

Elements of $Aut(G)$, when composed with elements of $Aut_i(G)$, \emph{partition} $Aut(G)$ into equally-sized \emph{cosets}, i.e., divides $Aut(G)$ into non-empty pair-wise disjoint subsets whose union is $Aut(G)$. The (right) coset of $Aut_i(G)$ in $Aut(G)$ containing permutation $\sigma$ is the set $\{\gamma \sigma | \gamma \in Aut_i(G)\}$. Choosing one element from each coset yields a set of \emph{coset representatives}. Each coset representative composed with $Aut_i(G)$ can generate the entire coset. 

Given $Aut(G)$, $i\sim j$ is defined for $i,j \in V$ (read $i$ and $j$ share the same \emph{orbit}), if and only if there exists symmetry $\gamma \in Aut(G)$ that maps $i$ to $j$, i.e., $\gamma i =j$. The $\sim$ operation imposes an \emph{equivalence} relation on $V$, which partitions $V$ into a so-called \emph{orbit partition}. The subsets of $V$ in the orbit partition are referred to as the \emph{orbits}. 

An \emph{ordered partition} $\pi=[W_1 \vert W_2 \vert \cdots \vert W_m]$ of $V$ is a partition of $V$ in which the order of subsets matters. The subsets $W_i$ are called the \emph{cells} of the partition. Ordered partition $\pi$ is \emph{unit} if $m=1$ (i.e., $W_1 = V$) and \emph{discrete} if $m=n$ (i.e., $|W_i|=1$ for $i=1,\cdots,n$). Ordered partition $\pi$ is \emph{equitable} with respect to graph $G$ if, for all $v_1, v_2 \in W_i$ ($1\le i\le m$), the number of neighbors of $v_1$ in $W_j$ ($1\le j\le m$) is equal to the number of neighbors of $v_2$ in $W_j$.

An \emph{ordered partition pair (OPP)} $\Pi$ is specified as 
\[
\Pi  = \left[ {\begin{array}{*{20}c}
   {\pi _T }  \\
   {\pi _B }  \\
\end{array}} \right] = \left[ {\begin{array}{*{20}c}
   {T_1 \left| {T_2 \left| { \cdots \left| {T_m } \right.} \right.} \right.} \hfill  \\
   {B_1 \left| {B_2 \left| { \cdots \left| {B_k } \right.} \right.} \right.} \hfill  \\
\end{array}} \right]
\]
with $\pi_T$ and $\pi_B$ referred to, respectively, as the top and bottom ordered partitions of $\Pi$. OPP $\Pi$ is \emph{isomorphic} if $m=k$ and $|T_i|=|B_i|$ for $i=1,\cdots,m$; otherwise it is \emph{non-isomorphic}. In other words, an OPP is isomorphic if its top and bottom partitions have the same number of cells, and corresponding cells have the same cardinality. An isomorphic OPP is \emph{matching} if its corresponding non-singleton cells are \emph{identical}. We refer to an OPP as discrete (resp. unit) if its top and bottom partitions are discrete (resp. unit). 

\section{Symmetry Finding vs. Canonical Labeling}
\label{sec:compare}

\begin{figure}[p]
\centering
\includegraphics[scale=0.45,keepaspectratio=true]{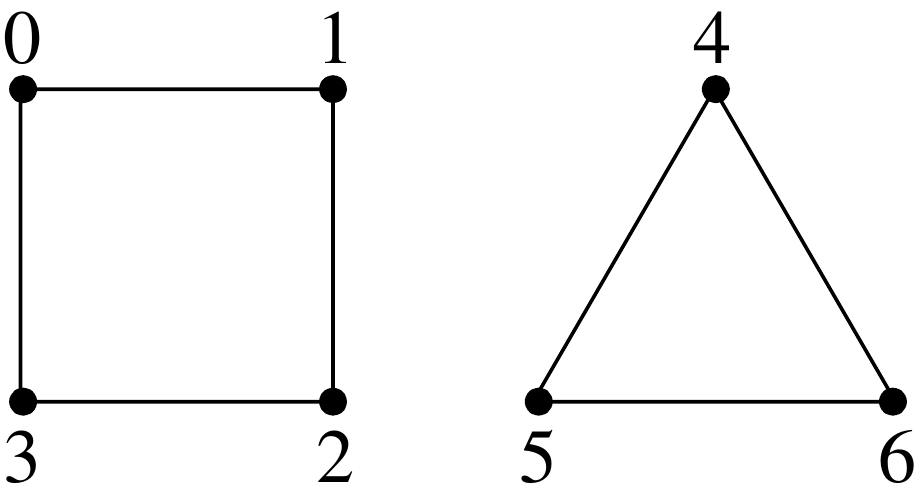}
\caption{A 7-vertex 7-edge graph with symmetry group of size 48.}
\label{fig:graph_example}
\vspace{19pt}
\centering
\includegraphics[scale=0.72,keepaspectratio=true]{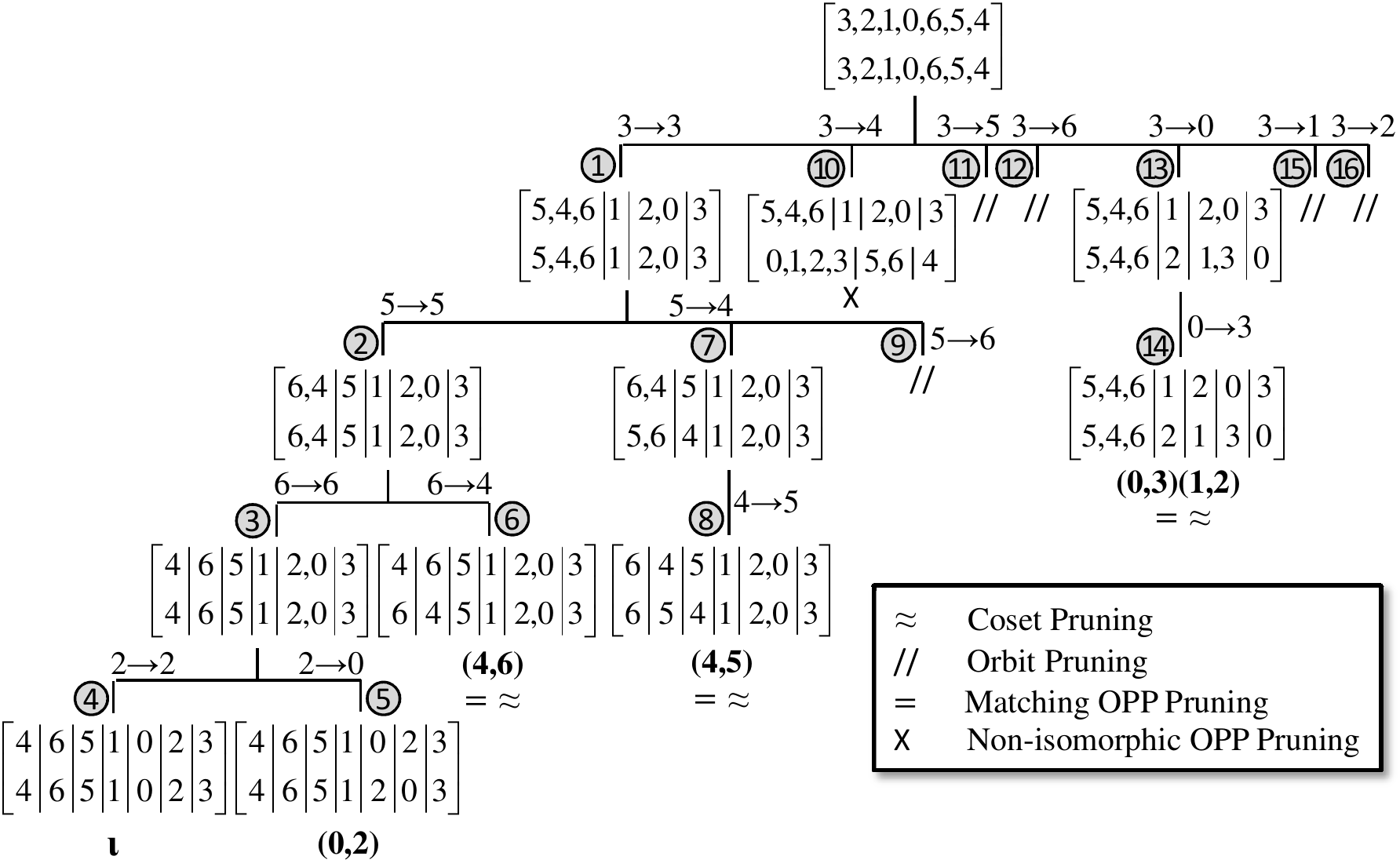}
\caption{Permutation tree constructed by \saucy{} 3.0 for the graph of Figure \ref{fig:graph_example}.}
\label{fig:saucy_tree}
\vspace{19pt}
\centering
\includegraphics[scale=0.72,keepaspectratio=true]{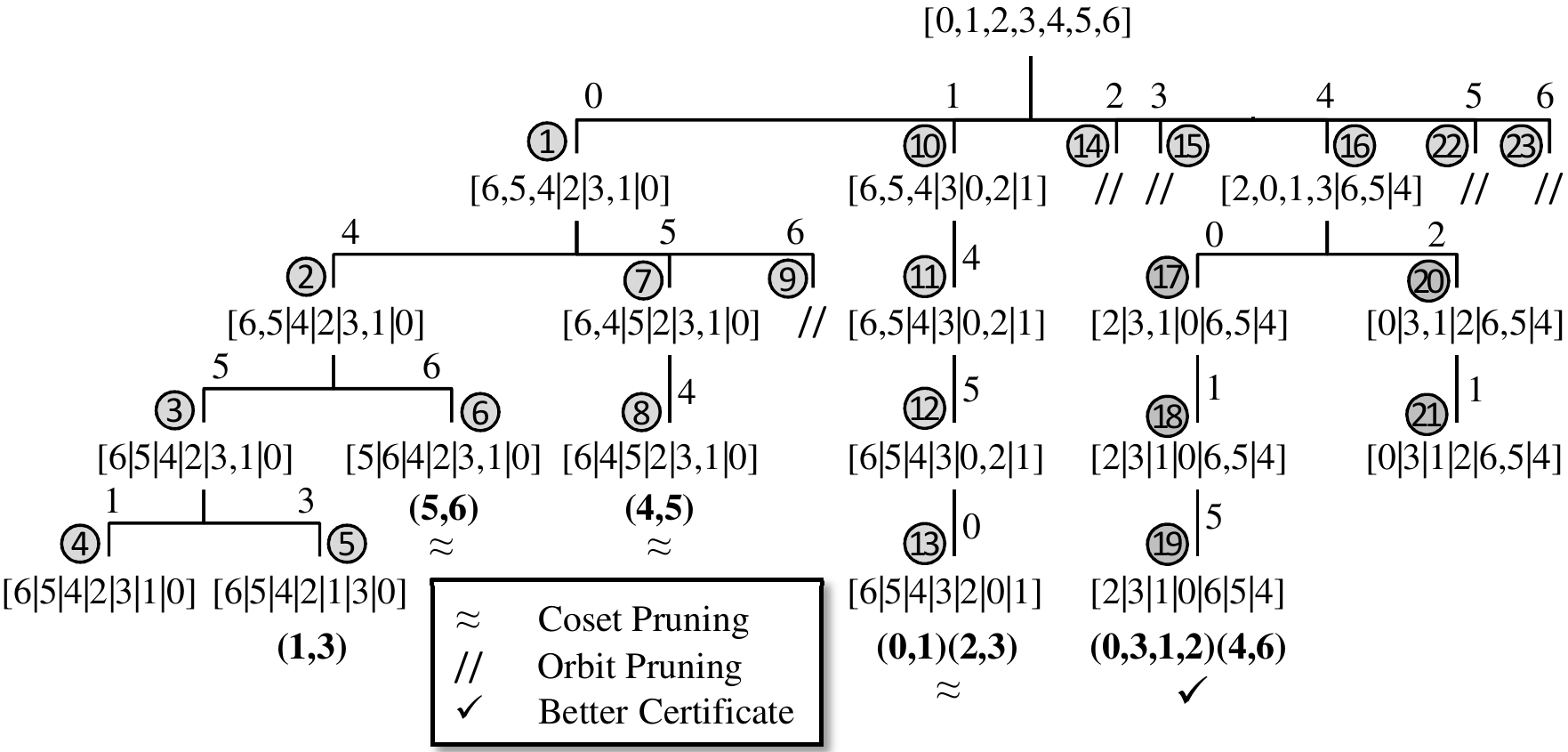}
\caption{Canonical labeling tree constructed by \bliss{} 0.72 for the graph of Figure \ref{fig:graph_example}.}
\label{fig:bliss_tree}
\end{figure}

In this section, we highlight the similarities and differences between the search for symmetries and a canonical labeling by focusing on the algorithms implemented in \saucy{} 3.0 and  \bliss{} 0.72. While we chose \bliss{} as a reference, our comparison can be extended to other \nauty{}-based canonical labeling tools. In the following subsections, we distinguish the search nodes of the trees constructed by \saucy{} and \bliss{}, explain what they represent, and show that the search trees used by these tools are fundamentally different. Furthermore, we discuss and compare the pruning techniques and branching mechanisms in \saucy{} and \bliss{}. We also point out an intrinsic limitation of the branching procedure in \bliss{}, and show that this limitation does not apply to \saucy{} search for automorphisms. To better understand and compare \saucy{} and \bliss{} baseline algorithms, an example graph (Figure \ref{fig:graph_example}) along with its search trees in \saucy{} (Figure \ref{fig:saucy_tree}) and \bliss{} (Figure \ref{fig:bliss_tree}) are provided. The nodes of the trees in Figures \ref{fig:saucy_tree} and \ref{fig:bliss_tree} are labeled in the order they are traversed by the \saucy{} and \bliss{} depth-first permutation search algorithms.

\subsection{Search Trees}
\label{subsec:trees}

The nodes of the search tree in \saucy{} are ordered partition pairs, each encoding a set of permutations. This set of permutations might be empty (non-isomorphic OPP), might have only one permutation (discrete OPP), or might consist of up to $n!$ permutations (unit OPP). In general, an isomorphic OPP
\begin{center}
$
\Pi  = \left[ {\left. {\begin{array}{*{20}c}
   {T_1 }  \\
   {B_1 }  \\
\end{array}} \right|\left. {\begin{array}{*{20}c}
   {T_2 }  \\
   {B_2 }  \\
\end{array}} \right|\left. {\begin{array}{*{20}c}
    \cdots   \\
    \cdots   \\
\end{array}} \right|\begin{array}{*{20}c}
   {T_m }  \\
   {B_m }  \\
\end{array}} \right]
$
\end{center}
represents
$
\prod\nolimits_{1 \le i \le n} {\left| {T_i } \right|} !
$
permutations. For example, the root of the tree in Figure \ref{fig:saucy_tree} is a unit OPP encoding all $7!=5040$ permutations on 7 elements, and the OPP at node $(7)$ is an isomorphic OPP representing the 4-element permutation set $\{(4,5),(4,5)(0,2),(4,6,5),(4,6,5)(0,2)\}$. 

In contrast, the nodes of the search tree in \bliss{} are single ordered partitions, each representing a (partial) labeling. 
A labeling in \bliss{} is obtained by renaming each vertex with the position of that vertex in the ordered partition. The ordering of vertices in the partition suggests a permutation that, when applied to the graph, produces the labeling encoded by that partition.  For example, at node (19) of Figure \ref{fig:bliss_tree}, vertices 0,1,2,3,4,5,6 are at indices 3,2,0,1,6,5,4, respectively, and hence, node (19) represents the labeling obtained by the permutation $(0,3,1,2)(4,6)$. To compare labelings, each node is associated with a \emph{certificate}. A certificate is a function that assigns a certain value to an ordered partition according to the graph's connection. Node certificates in \bliss{} are computed as follows. Given an equitable partition (returned by \emph{partition refinement}, see below), \bliss{} first makes a list of edges that connect singleton cells to other cells of the partition. For example, singleton cells $\{2\}$ and $\{0\}$ of the partition at node (1) of Figure \ref{fig:bliss_tree} are connected to cell $\{1,3\}$, and hence, the list of edges associated with node (1) is $\{\{2,1\}, \{2,3\}, \{0,1\}, \{0,3\}\}$. Then, \bliss{} generates the certificate by renaming each vertex in the list of edges with the index of that vertex in the partition. In our example, vertices 0,1,2,3 are at indices 6,5,3,4 of the partition at node (1), respectively, and hence, the certificate of node (1) is $\{\{3,5\}, \{3,4\}, \{6,5\}, \{6,4\}\}$. Two ordered partitions produce the same certificate if they are isomorphic to each other. For example, nodes (1) and (10) of Figure \ref{fig:bliss_tree} have the same certificate, but the certificate of node (1) is different from that of node (16).

To discard numerous impossible permutations and invalid labelings, \saucy{} and \bliss{} invoke \emph{partition refinement} after each branching decision. The goal of refinement is to propagate the constraints of the graph, i.e., the graph's vertex degrees, vertex colors, and edge relation, until the partition becomes equitable. Partition refinements in both \saucy{} and \bliss{} are adapted from \nauty{}; however, \saucy{}'s refinement benefits from \emph{simultaneous} comparison of the top and bottom partitions, a concept which is unique to the OPP representation of permutations. Simultaneous refinement allows \saucy{} to anticipate and avoid certain conflicts, which can lead to an exponential speed-up in run time \cite{Katebi-2011}.

The search for symmetries in \saucy{} starts by constructing a unit OPP at the root, and refining it. In the graph of Figure \ref{fig:graph_example}, all the vertices have the same color and degree, and hence, partition refinement does not distinguish any of the vertices. To explore the space of permutations, \saucy{} chooses a (target) vertex from a non-singleton (target) cell of the top partition and maps it to all the vertices of the corresponding cell of the bottom partition. For example, the target vertex at level 2 (nodes (2), (7), and (9)) of Figure \ref{fig:saucy_tree} is vertex 5, which is mapped to vertices 5, 4, and 6. Partition refinement is invoked after each mapping to prune away invalid permutations. The mapping procedure continues until the OPP becomes \emph{discrete}, \emph{matching}, or \emph{non-isomorphic} (e.g., nodes (5), (6), and (10) of Figure \ref{fig:saucy_tree}, respectively). A discrete or matching OPP represents a symmetry. A non-isomorphic OPP, however, indicates a conflict. The search ends when all possible mappings are exhausted.

The root of the canonical labeling tree in \bliss{} is a unit ordered partition which is initially refined. The depth-first traversal of permutation space starts by choosing a non-singleton cell, and individualizing all the vertices in that cell one at a time. For example, at level 2 (nodes (2), (7) and (9)) of Figure \ref{fig:bliss_tree}, all the vertices in the first non-singleton cell of the partition at node (1), i.e., vertices 4, 5, and 6, are individualized one after the other. Each individualization is followed by partition refinement to reflect the consequences of the branching decision. Individualization continues until the partition becomes discrete, i.e., the first leaf node is reached (node (4) of Figure \ref{fig:bliss_tree}). This leaf node is saved as a reference to compare certificates. A symmetry is found if another node during the search produces the same certificate as the first leaf node\footnote{Obtaining symmetries at non-leaf nodes will be discussed in Section \ref{subsec:pruning}.} (e.g., node (13) of Figure \ref{fig:bliss_tree}). The symmetry associated with such a node is the permutation that maps the partition at that node to the partition at the first leaf node. For example, the partition at node (13) of Figure \ref{fig:bliss_tree} encodes symmetry $(0,1)(2,3)$, since it can be obtained from the partition at node (4) by swapping vertex 0 with vertex 1 and vertex 2 with vertex 3. Furthermore, the canonical certificate is initialized to the certificate of the first leaf node, and is updated whenever a better certificate (based on any well-defined criterion, such as lexicographic ordering) is found during the search (e.g., node (19) of Figure \ref{fig:bliss_tree}). The canonical labeling of the graph is returned as the labeling of the node with the best certificate. 

\subsection{Pruning Techniques}
\label{subsec:pruning}

The search algorithms in \saucy{} and \bliss{} exploit two group-theoretical pruning mechanisms: \emph{coset pruning} and \emph{orbit pruning}. Coset pruning is based on the concept of coset representatives; one generator per coset is sufficient to generate all symmetries in the coset. For example, the symmetries found at node (8) of Figure \ref{fig:saucy_tree} and (13) of Figure \ref{fig:bliss_tree} are coset representatives of their corresponding subtrees rooted at node (7) and (10), and hence, those subtrees are coset pruned. Orbit pruning relies on orbit partition to eliminate redundant generators. For instance, node (9) of Figure \ref{fig:saucy_tree} is orbit pruned since vertices 5 and 6 share the same orbit. Similarly, node (9) of Figure \ref{fig:bliss_tree} is orbit pruned since vertices 4 and 6 share the same orbit. The algorithms for coset and orbit pruning follow similar implementations in \saucy{} and \bliss{}. 

To enable coset and orbit pruning, the left-most path of \saucy{} and \bliss{} search trees corresponds to a sequence of \emph{subgroup stabilizers} (a so-called \emph{subgroup decomposition}). In \saucy{}, stabilizers are maintained by mapping each vertex to itself (fixing each vertex) until the identity is reached. For example, the tree of Figure \ref{fig:saucy_tree} fixes vertices 3, 5, 6 and 2 to reach the identity at node (4). In \bliss{}, subgroup decomposition individualizes vertices one at a time until the partition is discrete. In the tree of Figure \ref{fig:bliss_tree}, stabilizer subgroups of 0, 4, 5 and 1 result in a discrete partition at node (4). 

In addition to these group-theoretic pruning techniques, the data structures in \saucy{} and \bliss{} allow additional pruning mechanisms. One such pruning mechanism in \saucy{} is \emph{non-isomorphic OPP pruning}. A non-isomorphic OPP contains permutations that do not form any symmetry. Such an OPP might be returned by \saucy{}'s partition refinement when a conflict is detected. For instance, the OPP at node (10) of Figure \ref{fig:saucy_tree} is non-isomorphic, which indicates that the mapping of 3 to 4 is a conflict. Similarly, \bliss{} identifies futile branches of the search by comparing the certificates of search nodes. Specifically, \bliss{} prunes a subtree if the certificate of the root of the subtree 1) does not match the certificate of the node on the left-most path of the tree at that level (i.e., the subtree does not yield any symmetry), and 2) is not better than the current best certificate (i.e, the subtree does not include the canonical labeling). For example, node (16) of Figure \ref{fig:bliss_tree} produces a different certificate than node (1), but the partial certificate associated with node (16) is better than the current best certificate (i.e, the certificate of node (4)), and hence, the subtree rooted at node (16) is explored. 

Another OPP-based pruning in \saucy{} is \emph{matching OPP pruning}. Recall that a matching OPP is a non-discrete OPP in which corresponding non-singleton cells contain the same elements. The significance of this OPP is that it represents an early automorphism constructed by mapping the vertices of non-singleton cells identically. This automorphism can be returned as the coset representative of the current subtree, which exempts the search from exploring the remaining permutations in that subtree. For example, the OPPs at nodes (6), (8) and (14) of Figure \ref{fig:saucy_tree} are found matching, and are returned as the coset representatives of the subtrees rooted at nodes (6), (7) and (13), respectively. Until recently, no pruning mechanism in \bliss{} had the same effect as the matching OPP pruning. In fact, all symmetries in \bliss{} were found at leaf nodes. However, recent advances in \bliss{} algorithms (version 0.72) exploit \emph{component recursion} to enable early detection of symmetries without reaching the leaves. This is accomplished by comparing the ordered partitions at each level to the left-most ordered partition at the same level. For example, partitions (6) and (3) of Figure \ref{fig:bliss_tree} both contain an identical non-singleton cell $\{3,1\}$. This suggests that node (6) represents the symmetry $(5,6)$, since partition (6) can be obtained from partition (3) by swapping vertex 5 with vertex 6. Although matching OPP and component recursion both aim to find symmetries early up in the tree, they are conceptually two distinct mechanisms, and impact the search trees in different ways. 

The \bliss{} algorithms use additional heuristics to facilitate the search for a canonical labeling. For example, \bliss{} stores recently discovered symmetries to (partially) detect and prune fruitless symmetric branches of the search. It also uses a methodology to propagate conflicts beyond the most recent branching points, which helps it expedite automorphism search by pruning away subtrees that yield the same conflict. These two pruning techniques are not implemented in \saucy{} 3.0, but our on-going research is investigating their possible incorporation.

\subsection{Branching Decisions}

Branching heuristics highly affect the performance of combinatorial search algorithms, including symmetry detection and canonical labeling. In \saucy{}, branching is performed by choosing a target cell and a target vertex from the top partition. On the left-most tree path, \saucy{} chooses the first non-singleton cell as the target cell, and the first vertex in that cell as the target vertex (see nodes (1) to (4) of Figure \ref{fig:saucy_tree}). In the remaining parts of the tree, \saucy{} looks for swaps of vertices, i.e., whenever it maps vertex $v_1$ to vertex $v_2$, it tries to map $v_2$ to $v_1$ right after. Note that this is not always possible as partition refinement might preclude the mapping of $v_2$ to $v_1$. In that case, \saucy{} picks the first vertex of any non-singleton cell of the top partition which is not identical to its corresponding cell of the bottom partition. The vertex-swap heuristic can also be viewed as a mechanism to maximize the occurrence of matching OPPs. For example, node (13) of Figure \ref{fig:saucy_tree} maps 3 to 0, and right after, node (14) maps 0 to 3. This consequently results in a matching OPP at node (14), representing the symmetry $(0,3)(1,2)$. In practice, this heuristic is most effective when symmetry generators are sparse.

The branching procedure in \bliss{} consists of a \emph{cell-selector} function. Given graph $G$ and partition $\pi$, cell-selector function $S(G, \pi)$ returns a non-singleton cell of $\pi$ such that $S(G,\pi)^\gamma = S(G^\gamma,\pi^\gamma)$ for all $\gamma \in Aut(G)$. The cell selector's latter condition ensures that the search trees constructed for isomorphic graphs are also isomorphic. In implementation, \bliss{} picks the same sequence of cells in all the paths from the root to the leaves. For example, the search tree of Figure \ref{fig:bliss_tree} always individualizes the vertices in the first non-singleton cell of the partition. The default branching heuristic in \bliss{} selects the \emph{maximum nonuniformly joined} cell, i.e., the first non-singleton cell which is nonuniformly joined to the maximum number of cells (two cells are nonuniformly joined if the vertices in one cell have both neighbors and non-neighbors in the other cell). In the search tree of Figure \ref{fig:bliss_tree}, maximum nonuniformly joined cells happen to be the first non-singleton cells of the partition.

Considering the structures of the search trees in \saucy{} and \bliss{}, \saucy{}'s branching procedure does not have the limitations of \bliss{}'s cell-selector function, since it can choose any target cell and target vertex at each step of the search. This consideration raises the possibility of improving the branching heuristic in \saucy{}. As mentioned, the default vertex-swap heuristic is effective when the input graph produces sparse generators. Our experimental results show that this is usually the case when the input graph is large and sparse. For other graphs, however, the vertex-swap heuristic does not necessarily produce the best results. We plan to explore other branching heuristics, and desirably, seek a methodology that adapts the branching heuristic to the characteristics of the input graph in our future research. 

\section{New Canonical Labeling Procedure}
\label{sec:can_approach}

In previous sections, we pointed out that \saucy{} algorithms and data structures were optimized to solve the graph automorphism problem, whereas, \bliss{} routines are mainly focused on finding a canonical representation. In this section, we propose a novel approach that takes advantage of \saucy{}'s efficiency in finding graph symmetries to speed up the search for a canonical labeling. We show that once the symmetries are found, canonical labeling can be performed much faster using this information by pruning the canonical labeling tree.

\begin{figure}[t]
\centering
\includegraphics[scale=0.5,keepaspectratio=true]{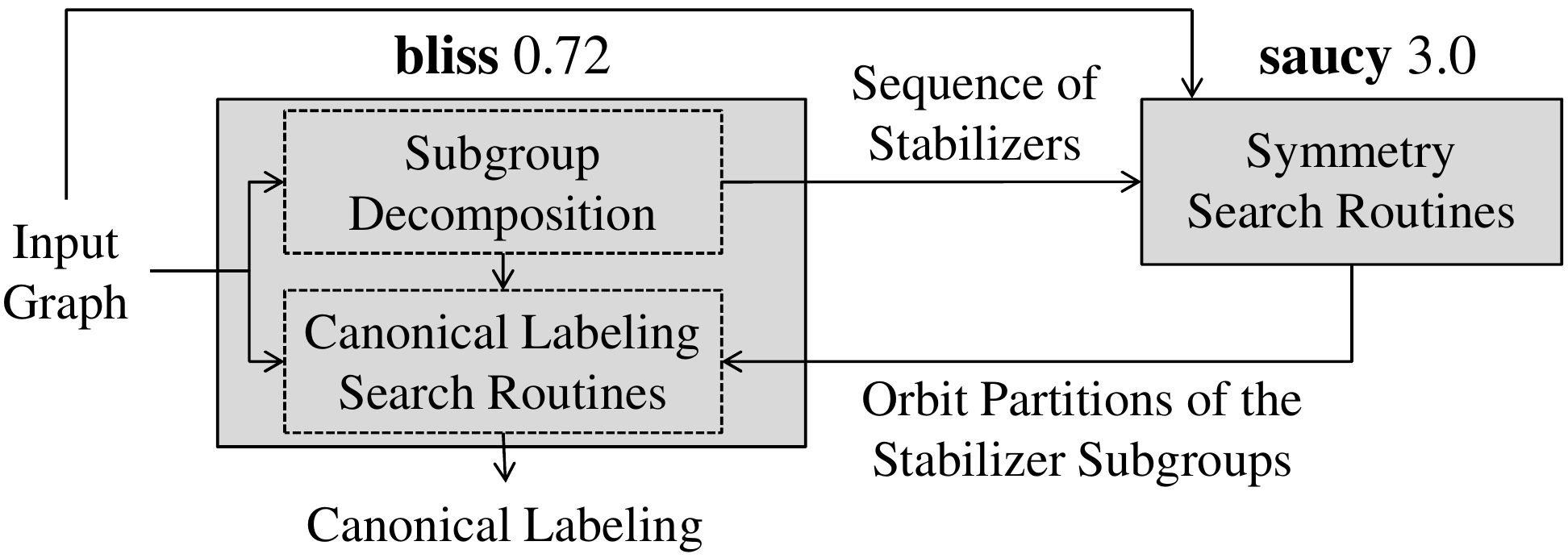}
\caption{Our proposed canonical labeling framework.}
\label{fig:can_flow}
\end{figure}

Our proposed graph canonicalization flow is depicted in Figure \ref{fig:can_flow}. It starts by launching \bliss{} to perform subgroup decomposition. Once decomposition is complete, it temporarily interrupts the search, passes the sequence of stabilizers obtained from subgroup decomposition to \saucy{}, and waits for \saucy{} to compute and pass back the graph's symmetry information. At the other end, \saucy{}'s decomposition routines use \bliss{}'s sequence of stabilizers to generate the  subgroups. In other words, the left-most path of the tree in \saucy{} is forced to match the one in \bliss{}. As \saucy{} looks for symmetries, it records the orbit partition at each level (i.e., the orbit partitions of the stabilizer subgroups). At the end of the search, it hands the computed orbit partitions over to \bliss{}. The canonical labeling algorithms in \bliss{} then resume the search, but incorporate two major modifications: 1) the level-by-level orbit partitions computed by \saucy{} are used to prune isomorphic subtrees, and 2) the search for symmetries is disabled in all expanded subtrees. Another way to say this is that a subtree that contains a symmetry will produce labelings that were previously examined, and hence, can be entirely pruned. On the other hand, a subtree that does not include any symmetry might lead to a better labeling (possibly, the canonical labeling), and hence, should be explored.

As elaborated above, our graph canonicalization approach divides the search into two phases: the search for symmetries and the search for a canonical labeling. In practice, this approach is effective when the input graph is highly symmetric, and the canonical labeling algorithms spend a lot of time looking for symmetries (instead of a canonical labeling). Our experimental results show that this logic applies when the input graph is large and sparse.

\section{Case Study}
\label{sec:case_study}

\begin{figure}[t!]
\centering
\includegraphics[scale=0.9,keepaspectratio=true]{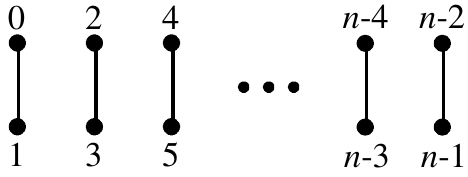}
\caption{An $n$-vertex $(n/2)$-edge graph with symmetry group size of $2^{n/2}\times (n/2)!$}
\label{fig:case_study_graph}
\centering
\includegraphics[scale=0.9,keepaspectratio=true]{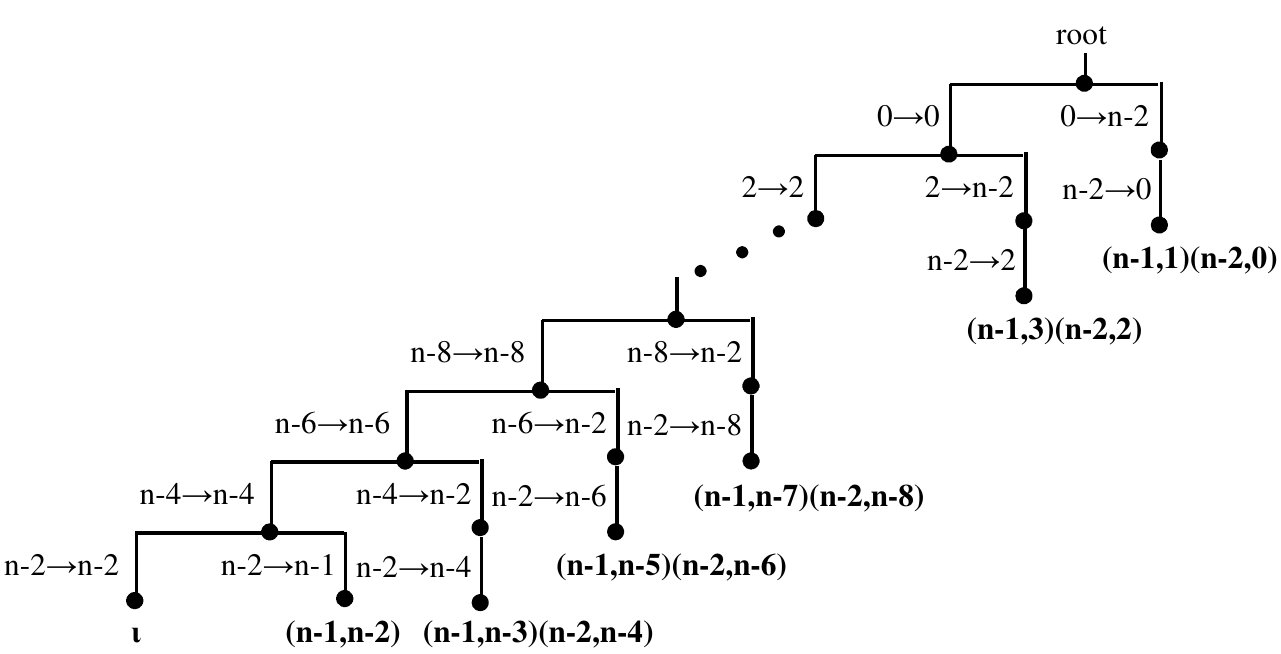}
\caption{Symmetry search tree constructed by \saucy{} for the graph of Figure \ref{fig:case_study_graph}.}
\label{fig:case_study_saucy}
\centering
\includegraphics[scale=0.9,keepaspectratio=true]{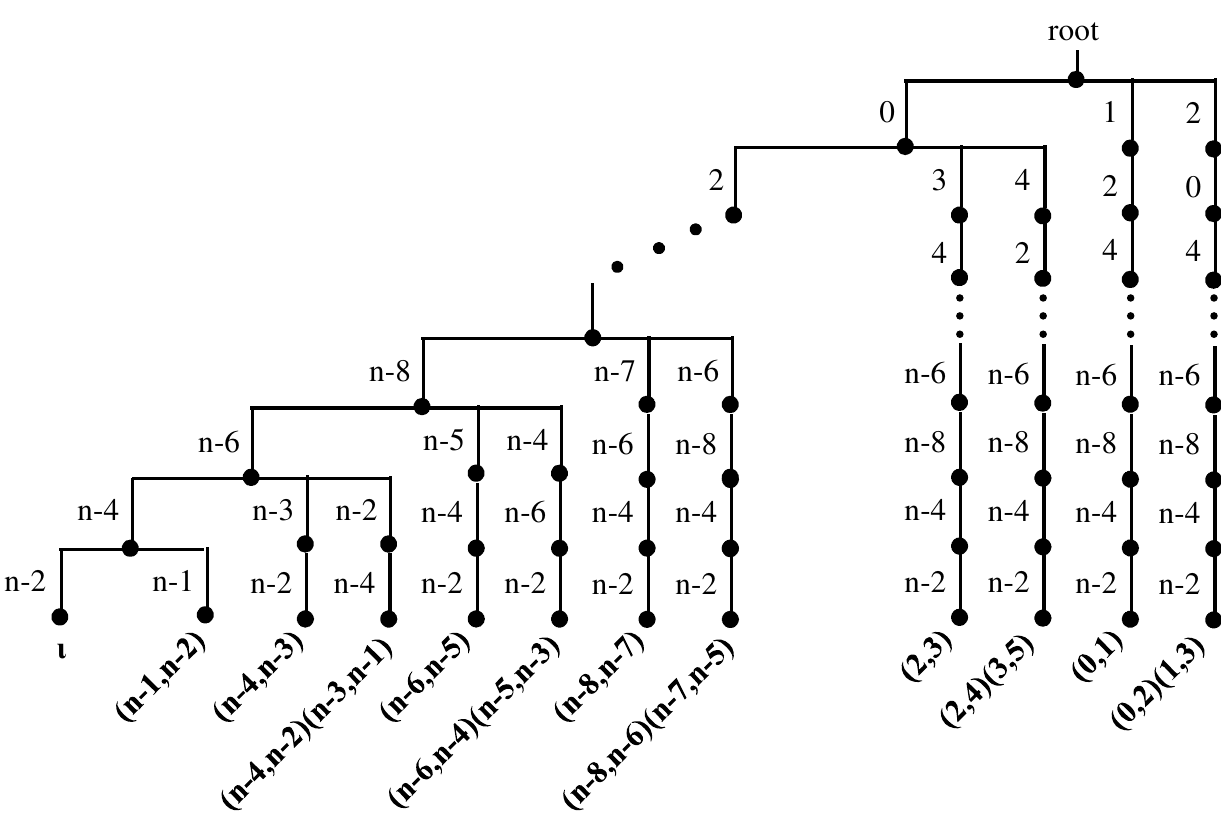}
\caption{Symmetry search tree constructed by \bliss{} for the graph of Figure \ref{fig:case_study_graph}.}
\label{fig:case_study_bliss}
\vspace{-10pt}
\end{figure}

This section analyzes and compares the run times of \saucy{} and \bliss{} in the search for the symmetries of an example graph shown in Figure \ref{fig:case_study_graph}. This graph has $n$ vertices, $n/2$ edges, average degree of 1, and the symmetry group size of $2^{n/2}(n/2)!$ The search trees generated by \saucy{} and \bliss{} for this graph are demonstrated in Figures \ref{fig:case_study_saucy} and \ref{fig:case_study_bliss}, respectively. The black nodes in these two trees correspond to OPPs/permutations. It is evident that \saucy{} explores fewer nodes than \bliss{}, as it finds symmetries up in the tree without reaching the leaves. A detailed analysis of run time complexity of \saucy{} and \bliss{} for this graph is presented next.

The \saucy{} search tree shown in Figure \ref{fig:case_study_saucy} produces $n/2$ levels after subgroup decomposition. The number of OPPs explored by \saucy{} at level $l$ is 3 for $2 \le l \le (n/2)$, 2 for $l = 1$, and 1 for $l = 0$ (root of the tree). The summation of all explored nodes over $n/2$ levels is:
\begin{center}
$\sum_{l=2}^{n/2} 3 + 2 + 1 = \sum_{l=1}^{n/2} 3 = 3n/2$
\end{center}

The \bliss{} search tree shown in Figure \ref{fig:case_study_bliss} produces $n/2$ levels after subgroup decomposition. The number of permutations explored by \bliss{} at level $l$ is $n$ for $l = n/2$, and $2l+1$ for $0 \le l < n/2$. The summation of all explored nodes over $n/2$ levels is:
\begin{center}
$n+\sum_{l=0}^{n/2-1} (2l+1) = n + 2\sum_{l=0}^{n/2-1} l + n/2 = 3n/2 + (n/2-1)(n/2) = n^2/4+n$
\end{center}

The analysis above shows that \saucy{} takes $\Theta(n)$ time to find the symmetries of the graph of Figure \ref{fig:case_study_graph}, while \bliss{} takes $\Theta(n^2)$. The combination of \saucy{} and \bliss{} takes $\Theta(n)$ time to canonically label the graph, due to the fact that all the $n$ vertices of the graph share the same orbit, and hence, all the subtrees encountered during \bliss{} canonical labeling search can be skipped. Our analysis discussed here matches empirical data presented next. 

\section{Empirical Validation}

We tested the performance of our proposed canonical labeling approach on 432 very large sparse graphs drawn from a wide variety of application domains. Our experiments were conducted on a SUN workstation equipped with a 3GHz Intel Dual-Core CPU, a 6MB cache and an 8GB RAM, running the 64-bit version of Redhat Linux. A time-out of 1000 seconds was applied to all experiments. Table \ref{tab:benchmarks} lists the families of the graph benchmarks in our suite.
These families are divided into three categories:

$-$ \saucy{} benchmarks: this set contains 92 very large and very sparse graphs first assembled to test \saucy{}. This suite represents graphs from logic circuits and their physical layouts \cite{Velev-DAC02,ISPD2005}, internet routers \cite{Cheswick00,Govindan00}, and road networks in the US states and its territories \cite{USCensusBureau}.

$-$ SAT 2011 benchmarks: this set includes a subset of the SAT 2011 competition CNFs \cite{SATCompetition} that have more than 10,000 variables (of the 1200 CNF instances, 313 had more than 10,000 variables). The choice of 10,000 was based on our observation that the modeled graphs for CNF benchmarks with that many variables tend to be very large and sparse.

$-$ binary networks: this set consists of graph benchmarks proposed to test community-detection algorithms \cite{Lancichinetti-2009}. We generated 27 undirected and unweighted binary networks using the implementation of the procedure described in \cite{Lancichinetti-2009} (available at \cite{binnet}). We set the number of nodes to $\{1,...,9\}\times \{10^3, 10^4, 10^5\}$, but fixed the average degree to 2, the max degree to 4, the mixing parameter to 0.1, the minimum community size to 20, and the maximum community size to 50 in all instances.

\label{sec:results}
\begin{table}[t]
  \centering
  \caption{Benchmark families}
  \resizebox{\textwidth}{!} {
    \begin{tabular}{p{2.3cm}|C{0.8cm}|C{1.1cm}C{1.15cm}|C{1.5cm}C{1.6cm}|p{4cm}}
    \hline\noalign{\smallskip}
    Family & Inst. & \multicolumn{2}{c|}{Smallest Instance} & \multicolumn{2}{c|}{Largest Instance} & Description \\
           &           &         vertices       &     edges     &         vertices     &       edges    &             \\
    \hline\noalign{\smallskip}    
    {\tt circuit} \cite{Velev-DAC02,ISPD2005} & 33 & 3,575   & 14,625  & 4,406,950  & 8,731,076 & \saucy{} graphs drawn from\\
    {\tt router} \cite{Cheswick00,Govindan00} & 3 & 112,969  & 181,639   & 284,805 & 428,624  & place\&route, verification, \\
    {\tt roadnet} \cite{USCensusBureau} & 56 & 1,158   & 1,008 & 1,679,418 & 2,073,394    & routers and road networks \\
    \hline\noalign{\smallskip}    
    {\tt application} \cite{SATCompetition} & 232 & 21,369 & 290,248 & 32,813,545 & 65,487,132 & SAT 2011 application,\\
    {\tt crafted} \cite{SATCompetition} & 11 & 46,164 & 365548 & 776,820 & 3,575,337 & crafted and random CNFs\\
    {\tt random} \cite{SATCompetition}  & 70 & 93,000 & 5,375 & 310,000 & 680,000 & with $\#vars>10,000$\\
    \hline\noalign{\smallskip}
    {\tt binnet} \cite{Lancichinetti-2009,binnet} & 27  & 1,000 & 720 & 9,000,000  & 658,675  & binary networks \\
    \hline\noalign{\smallskip}
    \end{tabular}%
  }
  \label{tab:benchmarks}
  \vspace{-10pt}
\end{table}%

\begin{figure}[t]
\centering
\includegraphics[scale=0.55,keepaspectratio=true]{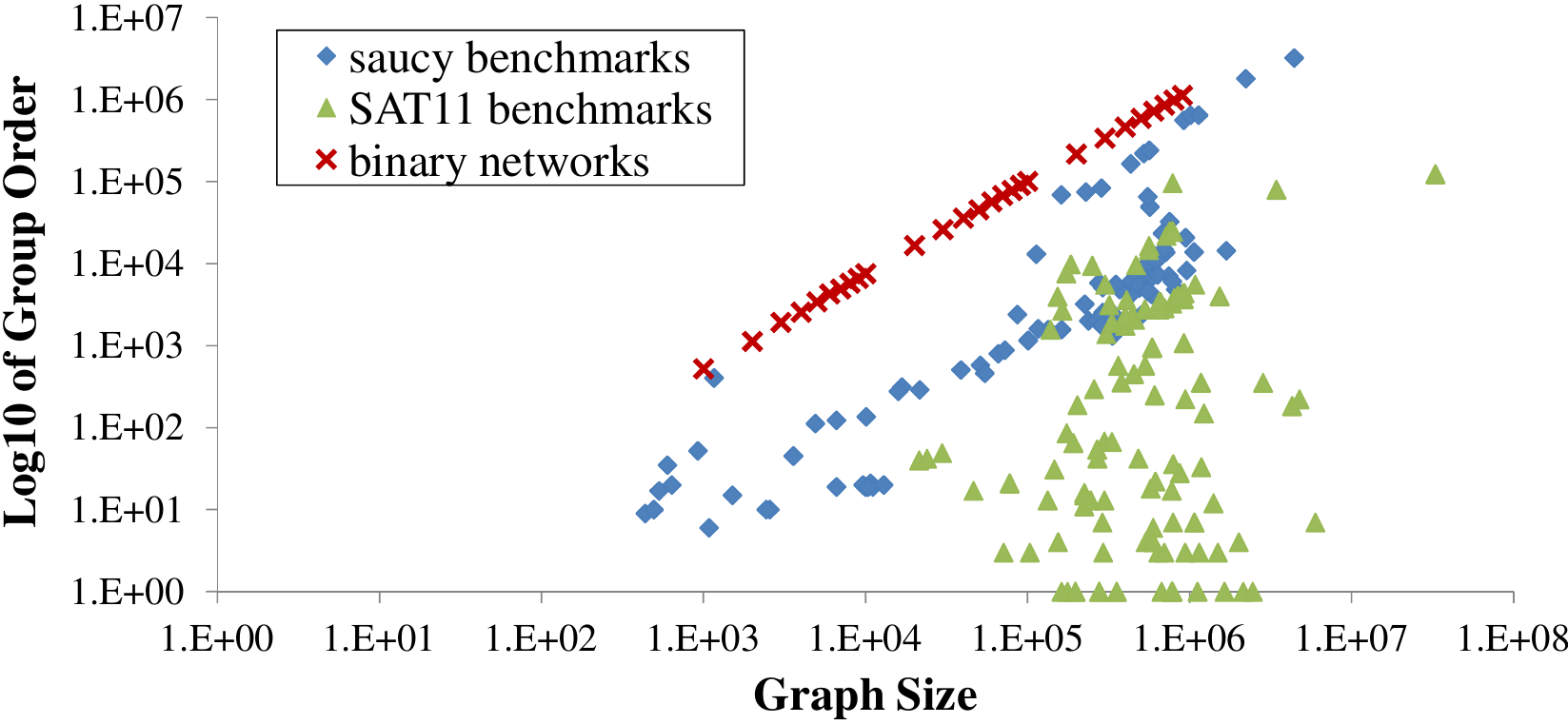}
\caption{Log-10 logarithm of symmetry group order as a function of graph size.}
\label{fig:group_size}
\vspace{-10pt}
\end{figure}

Our first set of experiments ran \saucy{} 3.0 \cite{saucy-3.0} on all graph benchmarks to find their symmetry group orders. Figure \ref{fig:group_size} depicts the base-10 logarithm of group order as a function of graph size. In total, 268 out of 432 benchmarks exhibited non-trivial symmetry. These 268 benchmarks included all \saucy{} graphs, all binary networks, and 149 out of 313 CNF instances.
The size of the largest symmetry group was $4\times 10^{3232782}$ 
and the smallest was 2. The results indicate that \emph{the majority of the graphs in our suite are highly symmetric}. Of the 268 graphs with at least one non-trivial symmetry, 203 ($75\%$) had group order of larger than $10^{10}$. 
 
\begin{figure}
  \centering
  \subfloat[]{\label{fig:bliss_sym_vs_can}\includegraphics[width=0.47\textwidth]{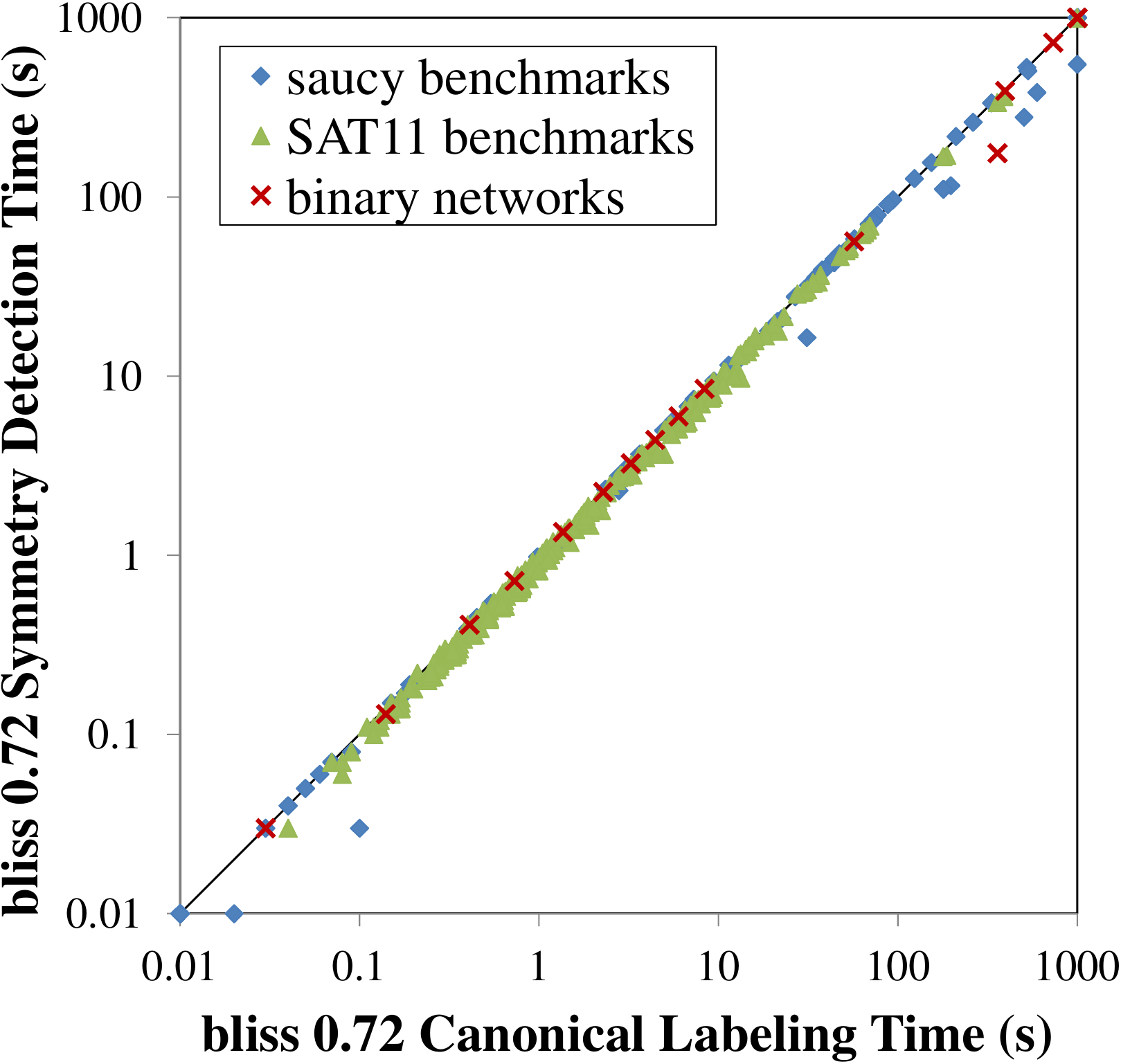}}                
  \hspace{8pt}
  \subfloat[]{\label{fig:saucy_vs_bliss}\includegraphics[width=0.47\textwidth]{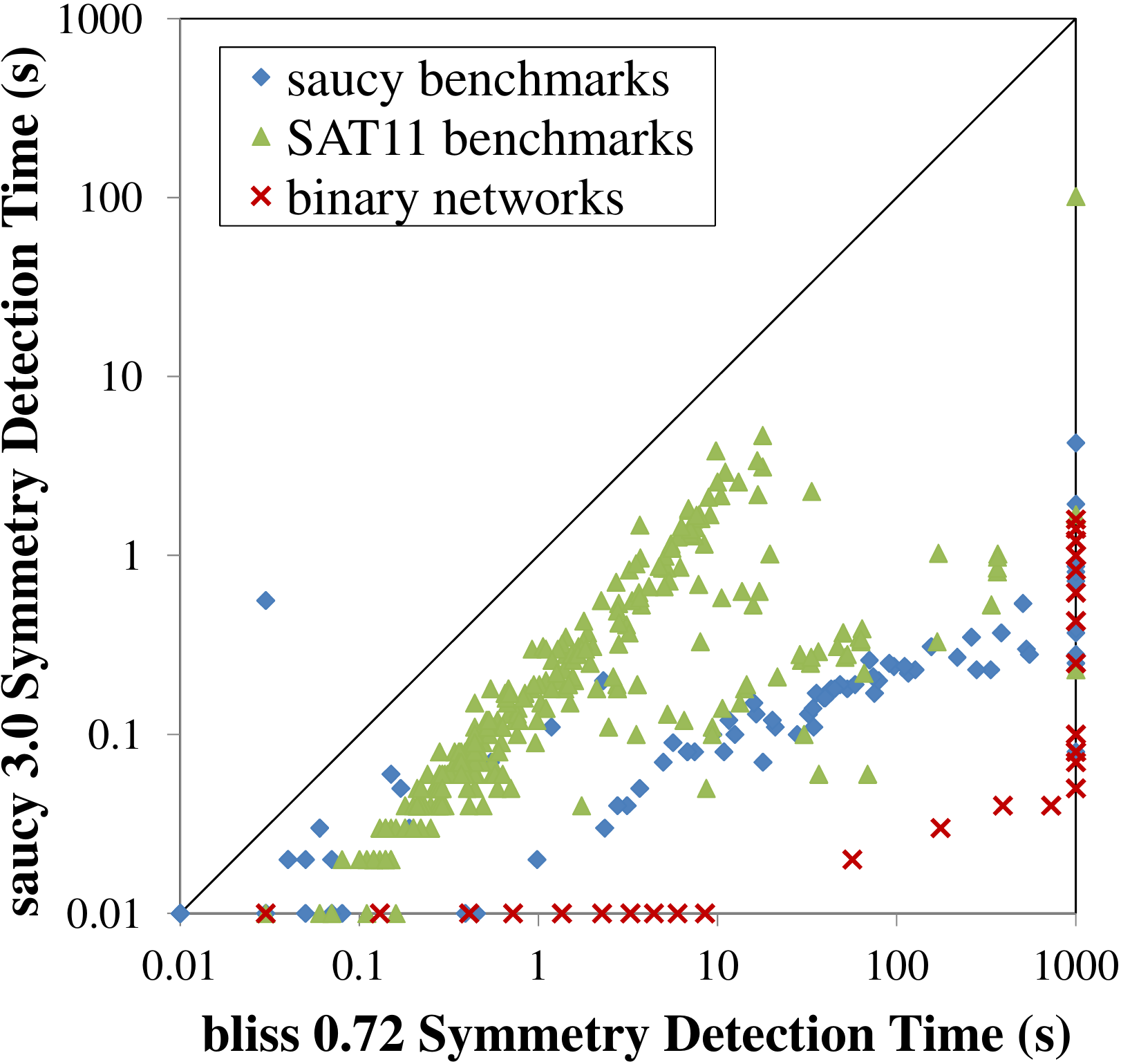}}                
  \caption{Run time comparison of (a) \bliss{} 0.72 symmetry detection versus canonical labeling, and (b) \saucy{} 3.0 versus \bliss{} 0.72 symmetry detection.}
  \label{fig:runtimes1}
\vspace{-10pt}
\end{figure}

To determine the amount of time that canonical labeling algorithms spend on finding automorphisms, we ran \bliss{} 0.72 \cite{bliss-0.72} under two configurations; once, to just search for symmetries, and once, to also look for a canonical labeling. Figure \ref{fig:bliss_sym_vs_can} depicts the results. It can be seen that \emph{the extra cost imposed by looking for a canonical labeling is negligible}. In other words, the canonical labeling routines spend most of their time searching for symmetries. 

To compare the performance of \saucy{} 3.0 and \bliss{} 0.72 in the search for automorphisms, we disabled the search for a canonical labeling in \bliss{}, and compared the results to those obtained from \saucy{}. This comparison is shown in Figure \ref{fig:saucy_vs_bliss}. The reported \saucy{} run times are obtained from its modified version which performs subgroup decomposition according to \bliss{}'s stabilizers sequence\footnote{For the run time comparison of the original \saucy{} 3.0 vs. \bliss{} 0.72, refer to \cite{Katebi-2011}.} (see Figure \ref{fig:can_flow}). It is evident that \emph{\saucy{} outperforms \bliss{} in the search for automorphisms}. There was only one benchmark from the \saucy{} suite that was processed by \saucy{} in 0.56 seconds, but was solved by \bliss{} in 0.03 seconds. Of the 432 total benchmarks, \saucy{} solved 431 in less than 5 seconds and processed one instance of SAT 2011 CNFs in 102 seconds, while \bliss{} timed out on 27, and solved the remaining in 727 seconds.
In general, \saucy{} benchmarks and binary networks seemed to be more challenging for \bliss{} than the SAT 2011 CNF instances. 



\begin{figure}[t!]
  \centering
  \subfloat[]{\label{fig:saucy_can_vs_bliss}\includegraphics[width=0.48\textwidth]{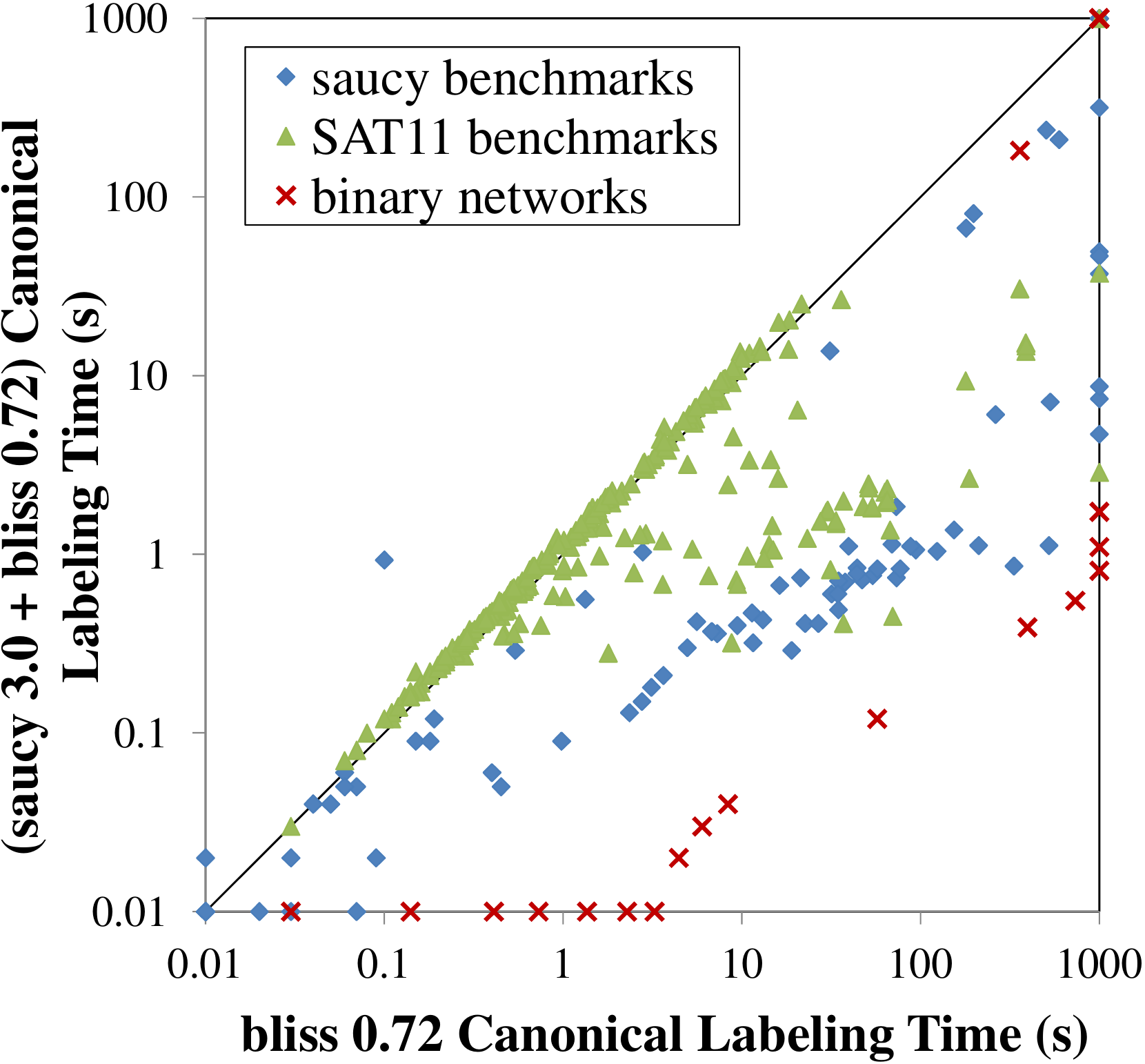}}                
  \hspace{8pt}
  \subfloat[]{\label{fig:saucy_can_vs_nauty}\includegraphics[width=0.48\textwidth]{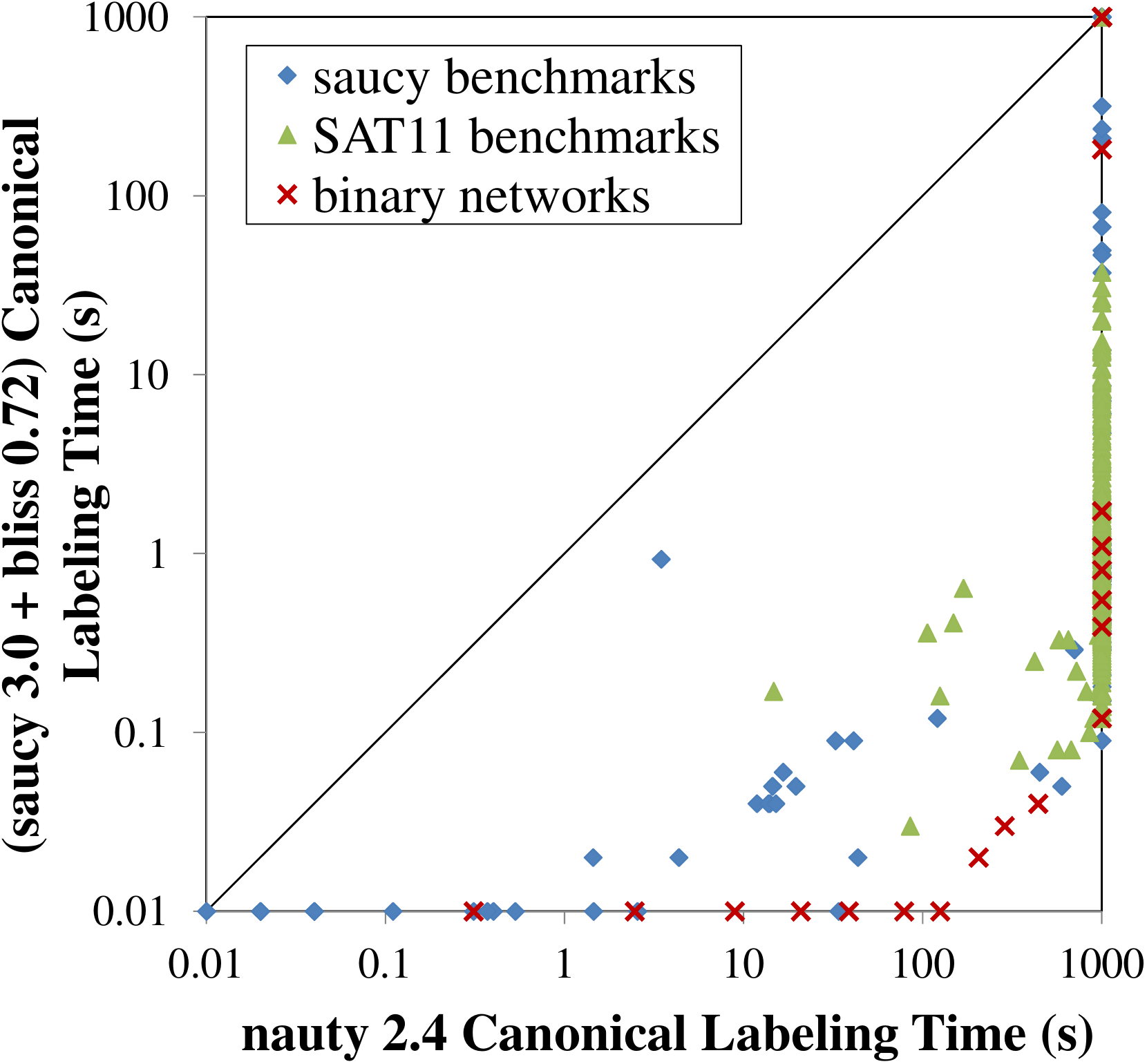}}
  \newline 
  \subfloat[]{\label{fig:saucy_can_vs_nishe}\includegraphics[width=0.48\textwidth]{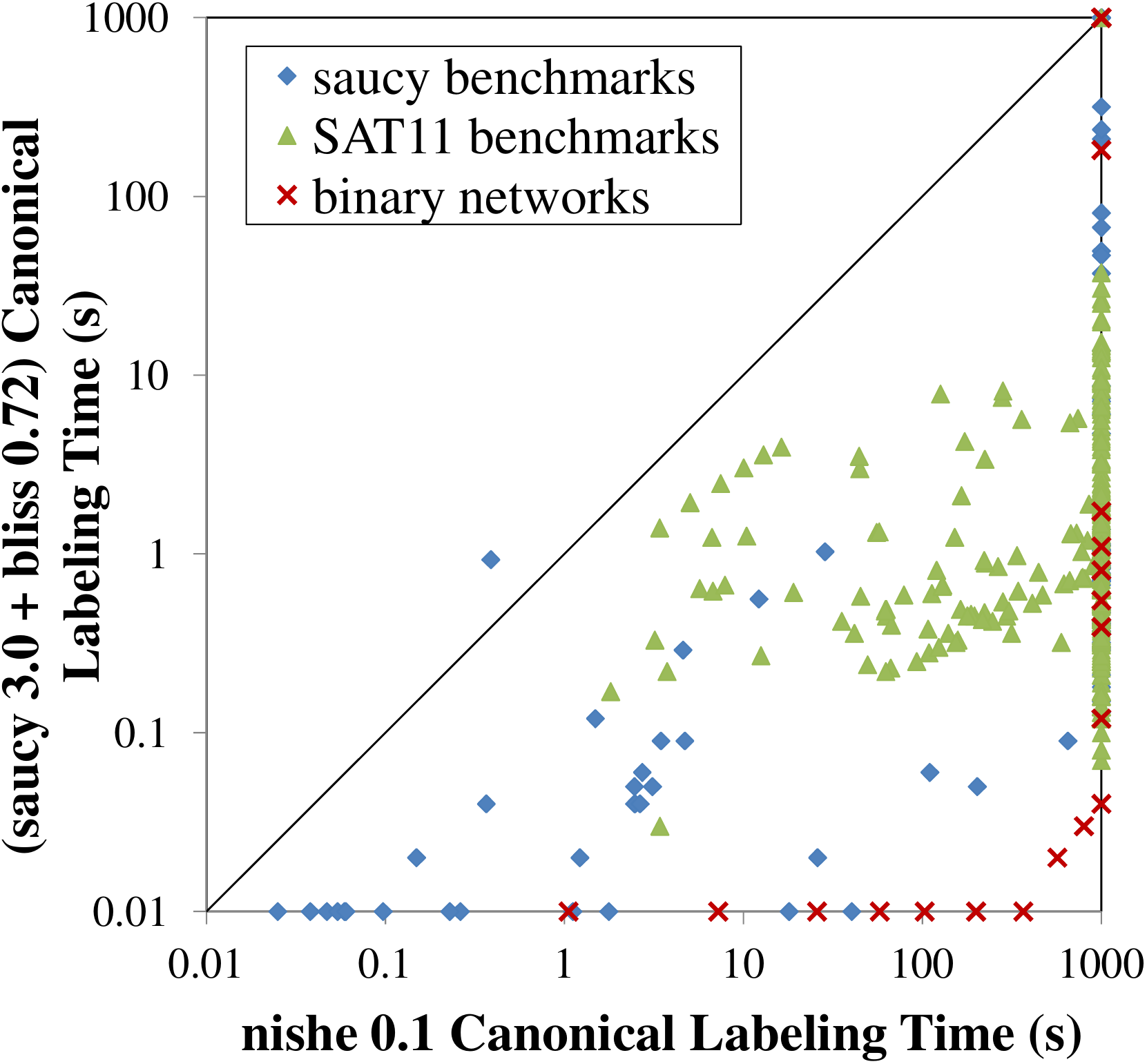}}                
  \hspace{8pt}
  \subfloat[]{\label{fig:saucy_can_vs_traces}\includegraphics[width=0.48\textwidth]{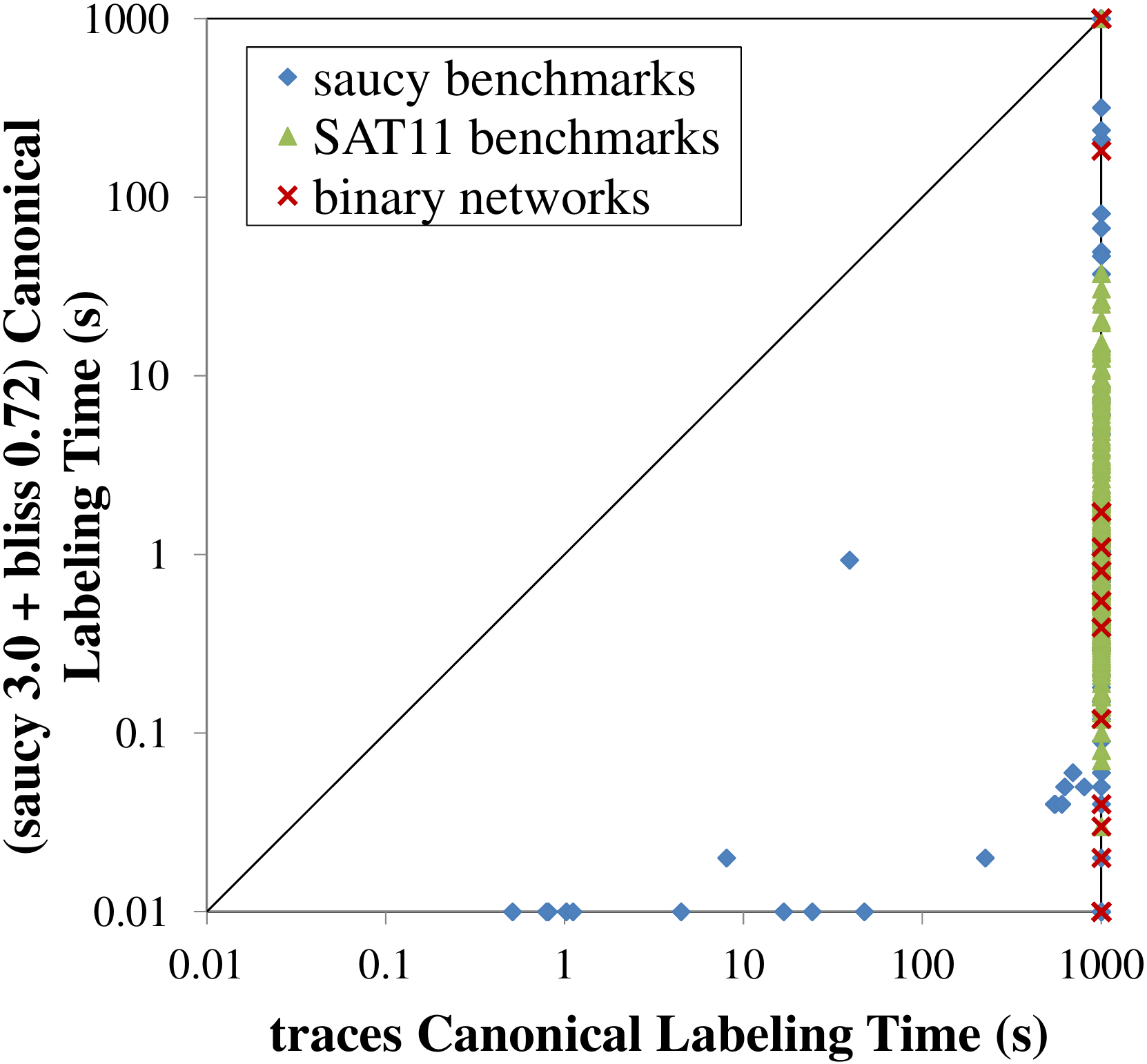}}                
  \caption{Run time comparison of the proposed canonical labeling approach in Figure \ref{fig:can_flow} versus (a) \bliss{} 0.72, (b) \nauty{} 2.4 (r2), (c) \nishe{} 0.1, and (d) \traces{} Nov09.}
  \label{fig:can_results}
\vspace{-10pt}
\end{figure}

To assess the performance of our proposed canonical labeling approach versus state-of-the-art canonical labelers, we compared the results of our approach to that obtained from \bliss{} 0.72 \cite{bliss-0.72}, \nauty{} 2.4 (r2) \cite{nauty-2.4}, \nishe{} 0.1 \cite{nishe-0.1}, and \traces{} Nov09 \cite{traces-Nov09}. Figure \ref{fig:can_results} depicts the results. These results clearly state that the combination of \saucy{} and \bliss{}, denoted by \saucy{}+\bliss{}, outperforms all the other four canonical labelers. Of the 432 total benchmarks, \saucy{}+\bliss{} solved 417, while \bliss{} solved 404, \nauty{} solved 58, \nishe{} solved 130, and \traces{} solved only 18. Furthermore, of the 432 benchmarks, 388 were solved by \saucy{}+\bliss{} in less than 10 seconds, while this number was reported to be 319 for \bliss{}, 18 for \nauty{}, 38 for \nishe{}, and 7 for \traces{}. Note that the 164 benchmarks that exhibited only one trivial symmetry did not benefit from our proposed canonical labeling approach. Nevertheless, the extra overhead imposed by those benchmarks was insignificant, as they were all processed by \saucy{} in less than 4 seconds. The detailed comparison between \saucy{}+\bliss{} and each of the four mentioned canonical labeling tools is presented next. 

Figure \ref{fig:saucy_can_vs_bliss} compares the run time of \saucy{}+\bliss{} to \bliss{}. In total, \bliss{} timed out on 28 benchmarks.
Of those 28, \saucy{}+\bliss{} managed to solve 13.
The remaining 404 benchmarks were solved by both \bliss{} and \saucy{}+\bliss{}. Of those 404 benchmarks, 137 experienced a speed-up by \saucy{}+\bliss{}, whereas, 223 went through a slow-down (slow-down for 164 benchmarks with symmetry group of size 1 was expected). The highest speed-up was 1334x, which was reported for the binary network with 50,000 vertices (run time was improved from 734 seconds to 0.55 seconds). The largest slow-down was 9x, which was reported for a \saucy{} graph.

Figure \ref{fig:saucy_can_vs_nauty} compares the run time of \saucy{}+\bliss{} to \nauty{}. In total, \nauty{} processed 58 benchmark,
but timed out or returned dynamic allocation failure on the remaining 374. All 58 benchmarks that were solved by \nauty{} were also solved by \saucy{}+\bliss{}. The largest reported run time from \nauty{} for those benchmarks was 956 seconds. This was while \saucy{}+\bliss{} processed all those benchmarks in less than a second. 

Figure \ref{fig:saucy_can_vs_nishe} compares the run time of \saucy{}+\bliss{} to \nishe{}. In total, \nishe{} failed to process 302 benchmarks, on which it either timed out, or returned a segmentation fault. 
Of these 302 benchmarks, \saucy{}+\bliss{} solved 287.
All the benchmarks that were solved by \nishe{} were also solved by \saucy{}+\bliss{}, but the run times of \saucy{}+\bliss{} were superior (speed-up of up to 36650x was reported). There was only one benchmark from the \saucy{} suite which was processed by \saucy{}+\bliss{} in 0.93 seconds, but was completed by \nishe{} in 0.38 seconds.

Figure \ref{fig:saucy_can_vs_traces} compares the run time of \saucy{}+\bliss{} to \traces{}. Of the 432 benchmarks, \traces{} only solved 18, all from the \saucy{} suite. The poor performance of \traces{} was due to the fact that it could not handle graphs with more than 18,000 vertices, and only 36 graphs in our suite (26 from \saucy{} benchmarks, and 10 from binary networks) exhibited less than 18,000 vertices. The 18 benchmarks that were processed by \traces{} were also processed by \saucy{}+\bliss{}, but a speed-up of up to 16025x was observed in \saucy{}+\bliss{} run times. 

In summary, the number of instances solved by each of the discussed canonical labeling tools suggests the following ordering of performance: \saucy{}+\bliss{} $>$ \bliss{} $>$ \nishe{} $>$ \nauty{} $>$ \traces{}. This ordering is obtained by testing each tool on a considerable number of large and sparse graphs. However, such an ordering is subject to a change if graphs with fewer vertices and higher edge concentration are used for benchmarking. 

\section{Conclusions}
\label{sec:conclusion}

Publications on graph automorphism and canonical labeling have typically focused on one of these problems and neglected the other. Canonical labeling algorithms produce symmetries as a byproduct, but are not as efficient as graph automorphism algorithms, which however, do not produce canonical labelings. This paper offers comparative analysis of relevant algorithms, highlighting the differences and exploring possible synergies. In particular, we show that canonical labeling algorithms can be more effective when symmetries are found in one dedicated pass and conveyed to these algorithms. We therefore develop an appropriate group-theoretic interface between \saucy{} --- the fastest symmetry finder --- and \bliss{} --- the fastest canonical labeler. Extensive empirical results convincingly demonstrate the benefits of our approach.

\bibliographystyle{plain}
\addcontentsline{toc}{section}{References}
\bibliography{Canonical}

\end{document}